\documentclass[aip,reprint,amsmath,amssymb]{revtex4-1}

\newif\ifpdf
        \ifx\pdfoutput\undefined
        \pdffalse 
        \else
        \pdfoutput=1 
        \pdftrue
        \fi
\ifpdf
   \usepackage[pdftex]{graphicx}
   \DeclareGraphicsExtensions{.pdf, .png, .jpg}
   \graphicspath{{./FIG/}}
   \usepackage{thumbpdf}
\else
   \usepackage{graphicx}
   \DeclareGraphicsExtensions{.eps}
   \graphicspath{{./FIG/}}
\fi

\usepackage{bm}
\usepackage{dcolumn}
\usepackage[yyyymmdd,hhmmss]{datetime}
\usepackage[colorlinks=true,urlcolor=blue,citecolor=blue,linkcolor=blue,breaklinks=true]{hyperref}
\usepackage{orcidlink}

\begin{document}


\title{When Theory Meets Experiment: What Does it Take to Accurately Predict $^1$H
NMR Dipolar Relaxation Rates in Neat Liquid Water from
Theory?}

\author{Dietmar Paschek\orcidlink{0000-0002-0342-324X}}
\email{dietmar.paschek@uni-rostock.de}
\affiliation{Institut f\"ur Chemie, Abteilung Physikalische und Theoretische Chemie, 
Universit\"at Rostock, Albert-Einstein-Str.~27, D-18059 Rostock, Germany}

\author{Johanna Busch\orcidlink{0000-0003-3784-0188}}
\affiliation{Institut f\"ur Chemie, Abteilung Physikalische und Theoretische Chemie, 
Universit\"at Rostock, Albert-Einstein-Str.~27, D-18059 Rostock, Germany}

\author{Angel Mary Chiramel Tony\orcidlink{0009-0003-7501-8483}}
\affiliation{Institut f\"ur Chemie, Abteilung Physikalische und Theoretische Chemie, 
Universit\"at Rostock, Albert-Einstein-Str.~27, D-18059 Rostock, Germany}

\author{Ralf Ludwig\orcidlink{0000-0002-8549-071X}}
\affiliation{Institut f\"ur Chemie, Abteilung Physikalische und Theoretische Chemie, 
Universit\"at Rostock, Albert-Einstein-Str.~27, D-18059 Rostock, Germany}

\author{Anne Strate\orcidlink{0000-0001-6621-8465}}
\affiliation{Institut f\"ur Chemie, Abteilung Physikalische und Theoretische Chemie, 
Universit\"at Rostock, Albert-Einstein-Str.~27, D-18059 Rostock, Germany}

\author{Nore Stolte\orcidlink{0000-0002-2892-2133}}
\affiliation{Lehrstuhl f\"ur Theoretische Chemie, Ruhr-Universit\"at Bochum, D-44780 Bochum, Germany}

\author{Harald Forbert\orcidlink{0000-0002-9992-2376}}
\affiliation{Center for Solvation Science ZEMOS, 
Ruhr-Universit\"at Bochum, D-44780 Bochum, Germany}

\author{Dominik Marx}
\affiliation{Lehrstuhl f\"ur Theoretische Chemie, 
Ruhr-Universit\"at Bochum, D-44780 Bochum, Germany}

\date{\today~at~\currenttime}

\begin{abstract}
In this contribution, we compute the 
$^1$H nuclear magnetic resonance (NMR) relaxation rate 
of liquid water at ambient conditions. 
We are
using structural and dynamical
information from Coupled Cluster Molecular Dynamics (CCMD) trajectories
generated at CCSD(T) electronic structure accuracy
while considering also nuclear quantum effects in addition to 
consulting information from X-ray and neutron scattering experiments. 
Our analysis is based on a recently presented
computational framework for
determining the frequency-dependent 
NMR dipole-dipole relaxation rate
of spin $1/2$ nuclei from Molecular Dynamics (MD) simulations, which allows for
an effective disentanglement of its
structural and dynamical contributions, and 
is including a
correction for finite-size effects inherent to MD simulations with
periodic boundary conditions. 
A close to perfect agreement with experimental relaxation data is
achieved
if structural and dynamical informations from  CCMD trajectories are considered
including a
re-balancing of the rotational and translational dynamics, 
according to the
product of the self-diffusion coefficient and
the reorientational correlation time of the H-H vector $D_0\times\tau_\mathrm{HH}$.
The simulations show 
that this
 balance is significantly altered when nuclear quantum effects are taken into account.
Our analysis suggests that the intermolecular and intramolecular
contribution to the $^1$H NMR relaxation rate of liquid water
are almost similar in magnitude, unlike
to what was predicted earlier from classical MD simulations.
\end{abstract}


\maketitle

\section{Introduction}

The nuclear magnetic resonance
(NMR) relaxation rate $R_1$ 
of $^1$H nuclei 
in liquid water has been 
measured in 1966 by K. Krynicki \cite{Krynicki_1966}
and shortly after that by E. v. Goldammer and M. D. Zeidler.\cite{Goldammer_1969}
It  was determined to be 
$R_1(\nu_\mathrm{H})\!=\!(0.280\pm0.003)\,\mbox{s}^{-1}$ 
at
a resonance frequency of $\nu_\mathrm{H}\!=\!28\,\mbox{MHz}$
at $298\,\mbox{K}$ 
and ambient pressure conditions.\cite{Krynicki_1966}
The question, we would like to answer in this contribution is:
How well do we understand and can hence predict this relaxation rate 
purely based on theory
almost 60 years later?

The primary mechanism for 
relaxation of spin $1/2$ nuclei in NMR spectroscopy is due to
the magnetic 
dipole-dipole interaction.\cite{Abragam1961,kowalewski_2013} 
Dipolar relaxation is sensitive to details of the intermolecular and intramolecular
structure and can
provide information about the molecular 
motions within a chemical system.\cite{Kruk_2011,Kruk_2012,honegger_2020} 
Interpreting experimental data, however, 
often requires analytical
models of the structure and dynamics to
accurately describe the underlying
 time correlation functions.\cite{Sholl_1981,Hwang_1975,overbeck_2020,overbeck_2021} 
Alternatively, also
Molecular Dynamics (MD) simulations can be used to study NMR relaxation phenomena, 
thus circumventing the need for analytical models. 
This advantage has been recognized early \cite{Westlund:1987,schnitker_1987}
and has provided detailed insights in the
NMR relaxation of ions in aqueous solution.\cite{carof_2014,carof_2015,carof_2016,chubak_2021,chubak_2023}
More recently, {\em ab initio} MD (AIMD) simulations  as
well as simulations
based on machine learning potentials
have come into focus as they can go beyond the 
limitations of classical molecular model potentials \cite{philips_2019,mazurek_2021} 
and can even account for nuclear quantum effects.\cite{Daru2022Coupled}
In addition, the electronic structure information from 
AIMD simulations also allows to determine quadrupolar NMR relaxation phenomena
for nuclei with spin $>1/2$.\cite{schmidt_2008,philips_2019a,philips_2020}

Dipolar NMR relaxation is due to the fluctuating fields resulting from the 
magnetic dipole-dipole
interaction 
between two spins, which can be formally divided into intermolecular
and intramolecular 
contributions.  
Here, the computation of
intermolecular relaxation rates needs to take
into account that
diffusion coefficients
obtained from MD simulations with periodic boundary conditions
 exhibit a 
 non-negligible system size dependence.\cite{yeh_2004,moultos_2016,busch_2023c,busch_2024a}
Moreover, the spectral densities at low frequencies cover motions
extending over 
large distance ranges.\cite{honegger_2021}
To deal with these problems, we have recently presented a 
computational framework to reliably
determine the frequency-dependent 
intermolecular
NMR dipole-dipole relaxation rate
from MD simulations.\cite{paschek_2024,paschek_2024cor} Our approach is based on a separation
of the intermolecular part into
a purely diffusion-based component, which is described
by the theory of Hwang and Freed \cite{Hwang_1975}
and a ``difference function''
computed
from MD simulations. It has been shown, that for long times this
``difference function''
quickly decays to zero and can thus be computed from rather
short MD simulation runs of modest size.\cite{paschek_2024,paschek_2024cor}
The influence of system-size dependent diffusion coefficients can be dealt with
by employing Yeh-Hummer \cite{yeh_2004,moultos_2016} corrected inter-diffusion 
coefficients for the Hwang and Freed model in combination with
a Yeh-Hummer-inspired scaling of the ``difference function''.\cite{paschek_2024,paschek_2024cor}
By construction,  this approach is capable of accurately predicting the 
correct low-frequency scaling behavior
of the intermolecular  NMR dipole-dipole relaxation rate.

We have recently computed the
frequency-dependent 
intermolecular and intramolecular dipolar NMR
relaxation rates of the
$^1$H nuclei in liquid water 
at $298\,\mbox{K}$
based on
simulations of
the TIP4P/2005 model for water.\cite{paschek_2024,paschek_2024cor} Our calculations
determine the relaxation rate to be 
$R_1(\nu_\mathrm{H})\!=\!0.3164\,\mbox{s}^{-1}$ 
at 
a resonance frequency of $\nu_\mathrm{H}\!=\!28\,\mbox{MHz}$,
which is $13\%$ above the experimental value.
In this contribution, we will show that this discrepancy
has to be mostly attributed to the intramolecular
relaxation rate
$R_{1,\mathrm{intra}}(\nu_\mathrm{H})$ 
and is to a large part the consequence of a misrepresentation of the
intramolecular H-H distance in the TIP4P/2005 model.
To account for  the influence of both the molecular structure
and dynamics
subject to nuclear quantum effects, we 
analyze
Coupled Cluster Molecular 
Dynamics (CCMD) 
trajectories from
path integral molecular dynamics simulations \cite{MarxHutter2009}
based on 
a high-dimensional neural network potential (HDNNP)~\cite{Behler2021Four}
that has been generated 
at CCSD(T) accuracy.\cite{Daru2022Coupled,stolte_2024}
In addition, we also consult structural information
from neutron scattering experiments reported by Soper,\cite{soper_2013}
as well as a recent re-evaluation of intramolecular H-H distances
based on a variety of experimental sources according to
Faux et al.\cite{faux_2021}.
We show that in addition to the intramolecular H-H distance
 a rebalancing of the reorienational dynamics
of the H-H vector in liquid water 
to match the balance between reorientational and translational dynamics
in CCMD simulations is essential for the agreement
with experimental $^1$H NMR relaxation rates.

\section{Theory}
\label{sec:theory}
\subsection{Dipolar NMR Relaxation and Correlations in the 
Structure and Dynamics of Molecular Liquids}

The dipolar relaxation of NMR active nuclei in a liquid is 
determined by their magnetic dipolar interaction
with all the nuclei in their surroundings and is subject to
the time-dependent spatial correlations in the liquid.
Here the NMR relaxation rate of nuclear
spins with $I = 1/2$ is dominated by the 
dipole-dipole interaction.\cite{Abragam1961} 
Hence the frequency-dependent relaxation rate is determined by the time dependence
of the magnetic dipole-dipole coupling.
For two like spins, it is described by \cite{Abragam1961,Westlund:1987}
\begin{eqnarray}\label{eq:relax}
R_1(\omega) & = &
\gamma^4\hbar^2I(I+1)
\left(\frac{\mu_0}{4\pi}\right)^2
\left\{
J(\omega) +4J(2\omega)
\right\}\;,
\end{eqnarray}
where $J(\omega)$
represents the spectral density of the time-dependent
nuclear dipolar interaction and
$\mu_0$ specifies the permeability of free space.
In case the of an isotropic fluid,
the spectral density is
given by \cite{Westlund:1987}
\begin{eqnarray}
\label{eq:jomega_1}
J(\omega) = \frac{2}{5}\;\mathrm{Re}\left\{ \int\limits_0^\infty 
G(t) \;e^{i\omega t} dt\right\}
\end{eqnarray}
where $G(t)$ denotes
the ``dipole-dipole correlation function''
which is available via
\cite{Westlund:1987,Odelius:1993}
\begin{eqnarray}\label{eq:dipolcor}
G(t) & = & \left< \sum_{j} r_{ij}^{-3}(0)\,r_{ij}^{-3}(t)
P_2\left[\,\cos \theta_{ij}(t) \right] \right>\;,
\end{eqnarray}
where
$\cos \theta_{ij}(t)$ is the cosine of the
angle between the connecting vectors $\vec{r}_{ij}$
joining spins $i$
and $j$ at time $0$ and at time $t$ while
$P_2[\ldots]$ represents the second Legendre polynomial.\cite{Westlund:1987}

By combining Equations \ref{eq:jomega_1} and \ref{eq:dipolcor},
the spectral density 
\begin{eqnarray}\label{eq:jomega}
J(\omega)&=&
\frac{2}{5}
\left<
  \sum_j r^{-6}_{ij}(0)
\right> 
\mathrm{Re}\left\{
\int\limits_0^\infty G^\mathrm{n}(t) \;e^{i\omega t} dt
\right\}
\end{eqnarray}
can be expressed as being composed of
a $r_{ij}^{-6}$ averaged 
constant
depending solely on the structure, and the Fourier-transform
of a normalized correlation function $G^\mathrm{n}(t)=G(t)/G(0)$, 
which is sensitive to the dynamical features
of the liquid.

For the case of 
$\omega\rightarrow 0$
we obtain a relaxation rate 
\begin{eqnarray}
R_{1}(0)
& = &
\gamma^4 \hbar^2
I(I+1)
\left(\frac{\mu_0}{4\pi}\right)^2
\cdot 2\int\limits_0^\infty
G(t) \,dt\;,
\end{eqnarray}
where the integral over the dipole-dipole correlation function 
\begin{eqnarray}\label{eq:ddcorel}
\int\limits_0^\infty G(t) \,dt & = & \left<
  \sum_j r^{-6}_{ij}(0)
\right> \tau_G\;
\end{eqnarray}
is the product of the $r_{ij}^{-6}$ averaged constant
and a correlation time $\tau_G$, which is the time-integral
of the normalized correlation function 
\begin{equation}
\tau_G=
\int\limits_0^\infty
G^\mathrm{n}(t) \,dt\;,
\end{equation}
characterizing the dynamical processes
within the liquid.

For convenience, one may divide the spins $j$ into different
classes according to whether they belong to the same molecule as
spin $i$, or not, thus arriving at an inter- and 
intramolecular contribution to the relaxation rate
\begin{eqnarray}
R_1(\omega) = R_{1,\rm inter}(\omega) + R_{1,\rm intra}(\omega) ,
\end{eqnarray}
which are related to corresponding 
intermolecular and 
intramolecular
dipole-dipole correlation functions
$G_\mathrm{intra}(t)$ and $G_\mathrm{inter}(t)$.
The  intramolecular contribution is  due
to molecular reorientations and conformational changes, while
the intermolecular contributions are largely affected by the translational
mobility (i.e. diffusion), but is also sensitive to local mutual reorientations,
librational motions,
and conformational changes of adjacent molecules.

\paragraph{Intermolecular Contributions:}

The structure of a liquid can be described by
the intermolecular site-site pair correlation functions $g_{ij}(r)$, 
denoting the probability of finding
a second atom of type $j$ in a distance $r$ from a reference site 
of type $i$ according to \cite{Egelstaff}
\begin{equation}
g_{ij}(r) = \frac{1}{N_i\,\rho_j} 
\left< 
\sum_{k=1}^{N_i}
\sum_{l=1}^{N_j}
\delta(\vec{r}-\vec{r}_{kl})
\right>\;,
\end{equation}
where $\rho_j$ is the number density of atoms
(or spins) of type $j$.
The pre-factor  of the intermolecular dipole-dipole correlation
function is hence related to the pair distribution function via
an $r^{-6}$-weighted integral over the pair correlation function
\begin{equation}\label{eq:r6gr}
\left<\sum_j r^{-6}_{ij}(0)\right> = \rho_j \,4\pi
\int\limits_0^\infty r^{-6} \; g_{ij}(r)\; r^2\,dr\;.
\end{equation}
The integral in Equation \ref{eq:r6gr} contains all
the structural correlations affecting the spin pairs
under consideration.
Averaged intermolecular distances between two spins
$\alpha$ and $\beta$ are thus represented by the integral
\begin{equation}
  \label{eq:dhh-1}
  I_{\alpha\beta}=4\pi\int\limits_0^\infty r^{-6}\,g_{\alpha\beta}(r)\,r^2 dr\;.
\end{equation}
Relating the structure of the liquid to a structureless hard-sphere fluid, 
the integral
$I_{\alpha\beta}$ can be also described by a ``distance of closest approach'' (DCA)
$d_{\alpha\beta}$, which represents an integral of the
same size, but over a step-like 
unstructured pair correlation function according to
\begin{equation}
  I_{\alpha\beta}=4\pi\int\limits_{ d_{\alpha\beta}}^\infty r^{-6}\cdot 1\cdot r^2 dr=
  \frac{4\pi}{3}\cdot\frac{1}{d_{\alpha\beta}^3}\;.
\end{equation}
Hence the DCA can be determined with
the knowledge of $I_{\alpha\beta}$ as 
\begin{equation}
  \label{eq:dhh-2}
  d_{\alpha\beta}=\left[\frac{4\pi}{3}\cdot\frac{1}{I_{\alpha\beta}}\right]^{1/3}\;.
\end{equation}
This DCA is conceptually 
identical to the distance used in the structureless hard-sphere diffusion model
as outlined by Freed and Hwang.\cite{Hwang_1975} 

To determine the $d_{\alpha\beta}$ in
this paper, the integral $I_{\alpha\beta}$ is evaluated via
\begin{equation}
  \label{eq:dhh-3}
  I_{\alpha\beta}\approx 4\pi\int\limits_0^{R_\mathrm{c}} r^{-6}\,g_{\alpha\beta}(r)\,r^2 dr
  + \frac{4\pi}{3} R_\mathrm{c}^{-3}
\end{equation}
by numerically integrating over the pair correlation function
up to a cutoff distance $R_\mathrm{c}$ and then  corrected by adding the term
$4\pi/(3R_\mathrm{c}^3)$ as long-range correction.

\paragraph{Intramolecular Contribution:}

Intramolecular correlations are computed directly by
summation over all involved spin pairs.
%
For the case of water, there remains only a single intramolecular 
dipole-dipole interaction.
By definition does a rigid water model such as TIP4P/2005 
possess only a fixed H-H distance. 
Therefore the intramolecular contribution
to the relaxation rate is dominantly based on 
the reorientational correlation function of
the intramolecular H-H vector
\begin{equation}
\label{eq:reorhh}
C_2^\mathrm{HH}(t)=\langle P_2[\vec{u}_\mathrm{HH}(0)\cdot\vec{u}_\mathrm{HH}(t)] \rangle\;,
\end{equation} 
where $\vec{u}_\mathrm{HH}$ is a unit vector oriented along
the intramolecular H-H axis
and $P_2[\ldots]$ indicates the second Legendre polynomial.

\subsection{Intermolecular Dipole-Dipole Relaxation: MD-Simulation and the 
Theory of Hwang and Freed}

We define the normalized  intermolecular dipole-dipole correlation function
as
\begin{equation}
G^\mathrm{n}_\mathrm{inter}(t)=G_\mathrm{inter}(t)/G_\mathrm{inter}(0)\;.
\end{equation}
Following the approach of Hwang and Freed \cite{Hwang_1975}, we can 
give an 
analytical expression  for the
dipole-dipole
correlation function  of two
diffusing particles with a DCA $d$,
reflecting boundary conditions at $r=d$, and
an inter-diffusion coefficient $D'$ using
\begin{equation}
\label{eq:g2model_u}
G^\mathrm{n,HF}_\mathrm{inter}(u)= \frac{54}{\pi}\;
\int\limits_0^\infty \frac{x^2\cdot e^{-u\cdot x^2}\;dx}{81+9x^2-2x^4+x^6}\;,
\end{equation}
where $u$ denotes a reduced timescale 
\begin{equation}
u\equiv \frac{D't}{d^2}\;,
\end{equation}
and $x$ represents a reduced inverse distance scale 
$x\equiv d/r$. 
For long
times, ultimately, 
$G^\mathrm{n,HF}_\mathrm{inter}(t)$ exhibits a $t^{-3/2}$ scaling behavior
and can be expressed using the reduced time-scale $u$ as \cite{Hwang_1975}
\begin{equation}
\label{eq:Gu32}
\lim_{u\rightarrow \infty}G^\mathrm{n,HF}_\mathrm{inter}(u)\approx
\frac{1}{6\sqrt{\pi} \cdot u^{3/2}}\;.
\end{equation}
The correct representation of the power-law behavior at long times 
is particularly 
important for properly describing 
the low-frequency limit of the corresponding
spectral density function 
$\lim_{\omega\rightarrow 0} J^\mathrm{HF}_\mathrm{inter}(\omega)\propto \sqrt{\omega}$.
Using the normalized dipole-dipole correlation function
of the Hwang and Freed theory according to
Equation \ref{eq:g2model_u}, the corresponding
intermolecular spectral density 
 is given by 
\begin{eqnarray}
\hspace*{-20pt}
J^\mathrm{HF}_\mathrm{inter}(\omega_u)&=& \frac{2}{5}
\left<
  \sum_j r^{-6}_{ij}(0)
\right> 
\cdot 
J^\mathrm{n,HF}_\mathrm{inter}(\omega_u)\;,
\end{eqnarray}
where $J^\mathrm{n,HF}_\mathrm{inter}(\omega_u)$ represents
the ``normalized'' Hwang-Freed spectral density, obtained
as a Fourier transformation of $G^\mathrm{n}_\mathrm{inter}(t)$  with
\begin{eqnarray}
\label{eq:HW_sdensity}
J^\mathrm{n,HF}_\mathrm{inter}(\omega_u)&=& 
\frac{54}{\pi}\cdot\frac{d^2}{D'}
\\ \nonumber
&& \times \int\limits_0^\infty 
\frac{dx}{(81+9x^2-2x^4+x^6)(1+\omega_u^2/x^4)}\;,
\end{eqnarray}
where $\omega_u\equiv\omega \cdot d^2/D'$ denotes a
reduced frequency scale, corresponding to the 
reduced time scale $u$.
From Equation \ref{eq:HW_sdensity} follows directly
the spectral density for $\omega_u\rightarrow 0$
\begin{eqnarray}
J^\mathrm{HF}_\mathrm{inter}(0) &=& \frac{2}{5}
\left<
  \sum_j r^{-6}_{ij}(0)
\right> 
\cdot \frac{4}{9}\cdot\frac{d^2}{D'}\;,
\end{eqnarray}
where 
\begin{equation}
\label{eq:taug}
\tau_\mathrm{G,HF}\!=\!J^\mathrm{n,HF}_\mathrm{inter}(0)\!=\!
\frac{4}{9}\cdot \frac{d^2}{D'}
\end{equation}
represents the 
intermolecular dipole-dipole
``correlation-time''
obtained as integral over the 
normalized dipole-dipole correlation function $G^\mathrm{n,HF}_\mathrm{inter}(t)$.

The Hwang and Freed theory outlined
above describes the behavior of random walkers 
sampled from an infinitely large system.
In computer simulations of condensed matter systems, however, 
we mostly deal with finite system sizes
using periodic boundary conditions,
thus limiting the volume from which the starting positions
are sampled.
We have recently shown that the problem can be hedged
by approximating
the effect caused by the limited sampling volumes
on the correlation function according to the Hwang and Freed model given
in Equation \ref{eq:g2model_u}, by employing
a nonzero lower boundary value 
$x_\mathrm{P}\!=\!\theta\cdot d/R$ with
$\theta\!\approx\!2.53$
for the
integral of Equation \ref{eq:g2model_u}, leading to
\begin{equation}
\label{eq:g2model_u_approx}
G^\mathrm{n,HF}_\mathrm{inter}(u,x_\mathrm{P})=
\frac{1}{A(x_\mathrm{P})}\;
\int\limits_{x_\mathrm{P}}^\infty \frac{x^2\cdot e^{-u\cdot x^2}\;dx}{81+9x^2-2x^4+x^6}\; ,
\end{equation}
recognizing that the variable $x$ is essentially representing an inverse distance. 
The parameter $\theta$ has been determined empirically to provide the best agreement
with data obtained
random walker simulations for various values of $R/d$.\cite{paschek_2024,paschek_2024cor}
Here the normalisation constant 
\begin{equation}
A(x_\mathrm{P})=\int\limits_{x_\mathrm{P}}^\infty \frac{x^2\;dx}{81+9x^2-2x^4+x^6}
\end{equation}
needs
to be computed by numerical integration, except for $A(x_\mathrm{P}\!=\!0)=\pi/54$.
The deviation of the approximate expression given by Equation \ref{eq:g2model_u_approx} from
Equation \ref{eq:g2model_u} can then be quantified by
\begin{equation}
\label{eq:scaling}
s(u,x_\mathrm{P})=\frac{G^\mathrm{n,HF}_\mathrm{inter}(u,x_\mathrm{P}\!=\!0)}{G^\mathrm{n,HF}_\mathrm{inter}(u,x_\mathrm{P})}\;,
\end{equation}
where the numerator represents Equation \ref{eq:g2model_u}. As shown
in Ref. \cite{paschek_2024,paschek_2024cor}, Equation \ref{eq:scaling} very well captures
the initial effect due to the limited sampling volumes over a broad
range of $R/d$ values that would correspond to liquid water simulations
ranging from 2000 to about 16000 water molecules in a cubic unit cell.
Since for longer times Equation \ref{eq:g2model_u} leads
to an overcorrection, we have introduced
of a time-limit $t_\mathrm{tr}$, characterizing up to which the correction could
be meaningfully applied. Realising that the corresponding timescale
is governed by the ratio of the radius $R$ (or the half box size $L/2$) and the 
inter-diffusion coefficient $D'$, we get
\begin{equation}
t_\mathrm{tr}=\frac{R^2}{4\pi D'}
=\frac{L^2}{16\pi D'}\,,
\end{equation}
which will consistently result in a time range where the scaling function
$s(u,x_\mathrm{P})\leq 1.25$.

\section{Methods}

We have performed classical MD 
simulations of liquid water using the TIP4P/2005 model \cite{abascal_2005},
which has been demonstrated to accurately describe the
properties of water compared to other
simple rigid nonpolarizable water models.\cite{vega_2011}
All simulations were carried out at $298\,\mbox{K}$ 
 under $NVT$ conditions using system sizes of 512, 1024, 2048, 4096,
and 8192 molecules. 
After sufficient equilibration, MD simulations of 1\,ns length each were performed
 using \textsc{Gromacs} 5.0.6.\cite{gromacs4,gromacs3}
The integration time step for all simulations was $2\,\mbox{fs}$.
The temperature of the simulated systems was controlled by employing the
Nos\'e-Hoover thermostat.\cite{Nose:1984,Hoover:1985}
More details about the simulations are given in Ref.~\cite{paschek_2024,paschek_2024cor}.
 
To  go beyond the accuracy of the TIP4P/2005 force field and understand the effect of quantum delocalization of nuclei,
we analyzed constant-volume ring polymer molecular dynamics (RPMD) simulations 
\cite{habershon_2013}
with 32 Trotter replica (``beads") of H$_2$O and D$_2$O at 298 K with 
their experimental densities 
at 1 atm \cite{Wagner2000IAPWS, Herrig2018Reference}, and computed structural properties, diffusion coefficients, $C_2(t)$ orientational correlation functions, and $\tau_2$ orientational correlation times.
The inclusion of nuclear quantum degrees of freedom has recently been demonstrated to be
essential for reproducing 
the structural and dynamical properties of liquid water 
based on very accurate 
interactions.\cite{Daru2022Coupled, yu_2022,Chen2023_Data-Efficient}
%
The RPMD simulations were performed with the Coupled Cluster Molecular Dynamics (CCMD) approach for liquid water \cite{Daru2022Coupled},
which employs a high-dimensional neural network potential (HDNNP) \cite{Behler2007Generalized, Behler2021Four} to simulate liquid water with CCSD(T) accuracy.
The analyzed trajectories were produced with CP2K \cite{CP2K, Hutter2014CP2K} with its path integral \cite{Brieuc2020Converged} and HDNNP \cite{Schran2018High} modules.
Individual RPMD trajectories were 100 ps long with a 0.25 fs time step and were initialized from configurations drawn at 10 ps intervals from path integral molecular dynamics simulations thermostatted with the path 
integral Langevin equation approach.\cite{Ceriotti2010Efficient}
Structural properties,
reorientational correlation functions $C_2(t)$, and correlation
times $\tau_2$ were obtained from RPMD trajectories with 256 molecules, with a total length of 5 and 11 ns for D$_2$O and H$_2$O, respectively.
Structural properties were extracted based on bead positions printed at 100 fs intervals.
$C_2(t)$ functions were obtained using bead positions \cite{Miller2005Quantum} at 1 fs intervals, and
we computed $\tau_2$ of the intramolecular H--H and D--D vectors, $\tau_2^{\mathrm{HH}}$ and $\tau_2^{\mathrm{DD}}$, by integrating $C_2^{\mathrm{HH}}(t)$ and $C_2^{\mathrm{DD}}(t)$ from the RPMD simulations from 0 to 40 ps following earlier work \cite{Wilkins2017Nuclear},
with error bars as described previously \cite{Daru2022Coupled}.
Self-diffusion coefficients were computed from individual simulations 
via a maximum entropy formalism \cite{Esser2018Tagging} based on 
molecular centroid positions at $2\,\mbox{fs}$ intervals, 
and system-size independent self-diffusion 
coefficients $D_0$ were obtained by extrapolating
results from cubic simulation cells containing 
64, 96, 128 and 256 molecules to the infinite box size.\cite{duenweg_1993, yeh_2004}
In total, we analyzed 5 and 11 ns of RPMD simulations at each box size for D$_2$O and H$_2$O, respectively.

Inter- and intramolecular dipole-dipole correlation functions 
obtained from classical MD simulation data
were computed using
our open-source software package ``MDorado'' which is available via GitHub (github.com/Paschek-Lab/MDorado).
Our open source 
software  ``FreeDRelax''  for 
performing finite-size cutoff-corrections to the intermolecular dipole-dipole correlation
functions and for
computing 
the inter- and intramolecular spectral densities and relaxation rates
discussed in section \ref{sec:theory}
is also available via GitHub (github.com/Paschek-Lab/FreeDRelax).

\section{Results and Discussion}

\subsection{Transport Properties}

Classical MD simulations of TIP4P/2005 water at 
$298\,\mbox{K}$ were carried out 
 at a density of
$0.99712\,\text{g}\,\text{cm}^{-3}$ 
corresponding to a pressure of $1\,\mbox{bar}$
for system sizes between 512 and 8192 molecules.
Data characterizing the simulations can be found in 
\tablename\ 1 of Ref. \cite{paschek_2024,paschek_2024cor}. 
The system-size
independent self-diffusion coefficient and viscosity 
of the TIP4P/2005 model have been recently re-evaluated
by us 
for the same conditions as 
used here based on the ``OrthoBoXY'' approach \cite{busch_2024b}
leading 
to a diffusion coefficient
of $D_0\!=\!(2.299\pm 0.005)\times 10^{-9}\,\mbox{m}^2\,\mbox{s}^{-1}$
and a viscosity of
$\eta\!=\!(0.860\pm 0.018)\,\mbox{mPa}\cdot\mbox{s}$.\cite{busch_2024b}
These data are perfectly consistent with the
simulations of Gonz\'ales and Abascal who report a
 viscosity
value of $0.855\,\mbox{mPa}\cdot\mbox{s}$
for TIP4P/2005 \cite{gonzales_2010},
and agree with the experimental
self-diffusion coefficient at $298\,\mbox{K}$ 
of $2.3\times 10^{-9}\,\text{m}^2\,\text{s}^{-1}$ 
by Krynicki \cite{krynicki_1978}and
$2.299\times 10^{-9}\,\text{m}^2\,\text{s}^{-1}$ 
by Holz and co-workers using
$^1$H PFG NMR experiments.\cite{holz_2000}
The system-size independent 
self-diffusion coefficients $D_0$ based
on CCMD simulations\cite{stolte_2024} for H$_2$O
with $(2.34\pm 0.09)\times 10^{-9}\,\mbox{m}^2\,\mbox{s}^{-1}$
and for D$_2$O
with $(1.86\pm 0.02)\times 10^{-9}\,\mbox{m}^2\,\mbox{s}^{-1}$
are also consistent 
with the experimental data given by Mills for H$_2$O and D$_2$O with
$2.299\times 10^{-9}\,\text{m}^2\,\text{s}^{-1}$, and
$1.872\times 10^{-9}\,\text{m}^2\,\text{s}^{-1}$
respectively.\cite{mills_1973}
The agreement of the CCMD simulations
with both H$_2$O and D$_2$O experimental self-diffusion
coefficients is suggesting that the chosen
ring polymer molecular dynamics (RPMD) 
simulation setup discussed in Ref. \cite{Daru2022Coupled}
is 
perfectly capable of capturing the
effect of quantum delocalization of nuclei on the 
dynamical properties of liquid water.\cite{Daru2022Coupled,stolte_2024}

\subsection{Intramolecular Relaxation}

To describe the long-time intramolecular dipole-dipole correlation functions
we do not employ a physics-based mechanistic model, but
the empricial Kohlrausch-Williams Watts (KWW)
function using
\begin{equation}
\label{eq:KWW}
G^\mathrm{n,KWW}_\mathrm{intra}(t) = A_\mathrm{K}\cdot\exp\left[-  \left(\frac{t}{\tau_\mathrm{K}}\right)^{\beta_\mathrm{K}}\right]\;.
\end{equation}
This empirical model has been fitted to the computed $G^\mathrm{n,MD}_\mathrm{intra}(t)$
over a time interval between $1\,\mbox{ps}$  and $100\,\mbox{ps}$.
Here, the KWW function can be regarded as representing a distribution of relaxation times, where the parameter 
$\beta_\mathrm{K}$ describes the width of this distribution. Note, that it is common practice to assume a distribution of relaxation times to 
describe the intramolecular contribution to the 
NMR relaxation in molecular liquids.\cite{roessler_2019} 
This phenomenon has been recently extensively analyzed by Asthagiri and
co-workers.\cite{asthagiri_2020,parambathu_2024}

\begin{figure}
        \includegraphics[width=0.4\textwidth]{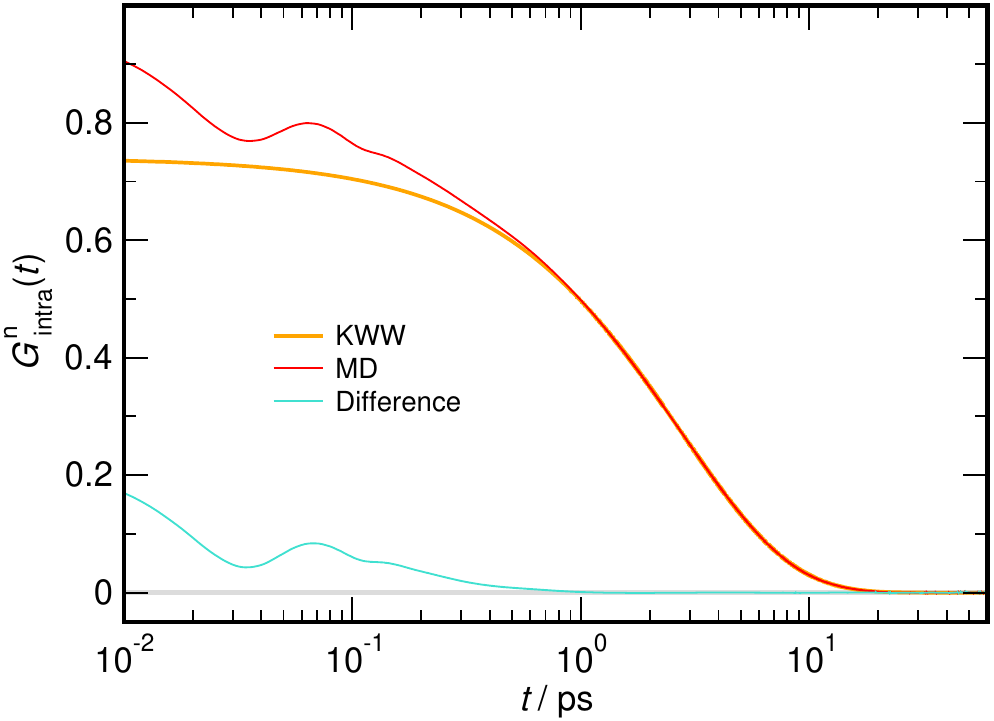}        
        \caption{\label{fig:g2_intra}
         Normalized intramolecular
         dipole-dipole correlation function
        $G^\mathrm{n}_\mathrm{intra}(t)$
        computed for water at $298\,\mbox{K}$.
        Intramolecular dynamics from CCMD simulations
        of H$_2$O at CCSD(T) accuracy which include nuclear quantum effects.
       Solid red line: $G^\mathrm{n,MD}_\mathrm{intra}(t)$. 
	    Solid orange line:
	     $G^\mathrm{n,KWW}_\mathrm{intra}(t)$ 
	     according to Equation \ref{eq:KWW}
	     using $A_\mathrm{K}\!=\!0.73995$, 
	     $\tau_\mathrm{K}\!=\!2.7622\,\mbox{ps}$, and
	     $\beta_\mathrm{K}\!=\!0.9057$, and
	     $\Delta G^\mathrm{n}_\mathrm{intra}(t)=G^\mathrm{n,MD}_\mathrm{intra}(t)-G^\mathrm{n,KWW}_\mathrm{intra}(t)$.
	    }
\end{figure}

In \figurename\ \ref{fig:g2_intra}
both functions are plotted 
from data of the CCMD simulations of H$_2$O.
A log-linear representation of the data, including the
difference function
\begin{equation}
\Delta G^\mathrm{n}_\mathrm{intra}(t)=G^\mathrm{n,MD}_\mathrm{intra}(t)-G^\mathrm{n,KWW}_\mathrm{intra}(t)
\end{equation}
is shown in \figurename\ \ref{fig:g2_intra}.
Significant differences between 
$G^\mathrm{n,MD}_\mathrm{intra}(t)$ and $G^\mathrm{n,KWW}_\mathrm{intra}(t)$ are restricted to
a time-interval $t\leq 1 \,\mbox{ps}$.
Hence
the intramolecular
correlation time $\tau_\mathrm{G}$ was computed 
as an integral over $G^\mathrm{n,MD}_\mathrm{intra}(t)$, 
which can be splitted into two terms
according to
\begin{equation}
\label{eq:deltagintra}
\tau_\mathrm{G} = \tau_\mathrm{G,KWW} + \Delta\tau_\mathrm{G}
\end{equation}
with 
\begin{equation}
\label{eq:tauKWW}
\tau_\mathrm{G,KWW}\!=\!A_\mathrm{K}\tau_\mathrm{K}\beta^{-1}_\mathrm{K}\Gamma(\beta^{-1}_\mathrm{K})\;,
\end{equation}
where $\Gamma(\ldots)$ represents the Gamma-function.

To compute the intramolecular
spectral density from MD simulation, we use
\begin{equation}
J^\mathrm{n,MD}_\mathrm{intra}(\omega)
=
J^\mathrm{n,KWW}_\mathrm{intra}(\omega)
+
\Delta J^\mathrm{n}_\mathrm{intra}(\omega)
\end{equation}
with
\begin{equation}
\label{eq:deltaj_md_reor}
\Delta J^\mathrm{n}_\mathrm{intra}(\omega)
\approx
\int\limits_0^{t^*} 
\Delta G^\mathrm{n}_\mathrm{intra}(t)
\cos (\omega t) \;dt\;.
\end{equation}
Here the computation of the integral
in Equation \ref{eq:deltaj_md_reor}
is as well performed numerically,
employing the trapezoidal rule
up to a time $t^*=5\mbox{ps}$, where
both functions $G^\mathrm{n,MD}_\mathrm{intra}(t)$
and $G^\mathrm{n,KWW}_\mathrm{intra}(t)$ become effectively indistinguishable.
The Fourier-transform of $G^\mathrm{n,KWW}_\mathrm{intra}(t)$, defined in
Equation \ref{eq:KWW}, $J^\mathrm{n,KWW}_\mathrm{intra}(\omega)$, needs, however, to be computed numerically, 
due to the lack of an analytical Fourier-transform
equivalent of the KWW function. 
To compute $J^\mathrm{n,KWW}_\mathrm{intra}(\omega)$ properly,
we have tested the convergence 
of the numerical cosine-transform evaluation by comparing it to the
limiting value for $\omega\rightarrow 0$.
The intramolecular
$^1$H NMR relaxation rate
following Equation \ref{eq:relax}
is finally computed via
\begin{eqnarray}
\label{eq:R1_intra}
R_{1,\mathrm{intra}}(\omega)
&=&
\gamma_\mathrm{H}^4\hbar^2 \cdot\frac{3}{4}\cdot\left(\frac{\mu_0}{4\pi}\right)^2\cdot
\frac{2}{5}\cdot
\frac{1}{r_\mathrm{HH}^6}\times\\
&&\left\{
J^\mathrm{n,MD}_\mathrm{intra}(\omega)
+
4\,J^\mathrm{n,MD}_\mathrm{intra}(2\omega)
\right\}\;,
\nonumber
\end{eqnarray}
where $r_\mathrm{HH}$ represents the properly weighted intramolecular
H-H distance.
\begin{figure}
        \includegraphics[width=0.4\textwidth]{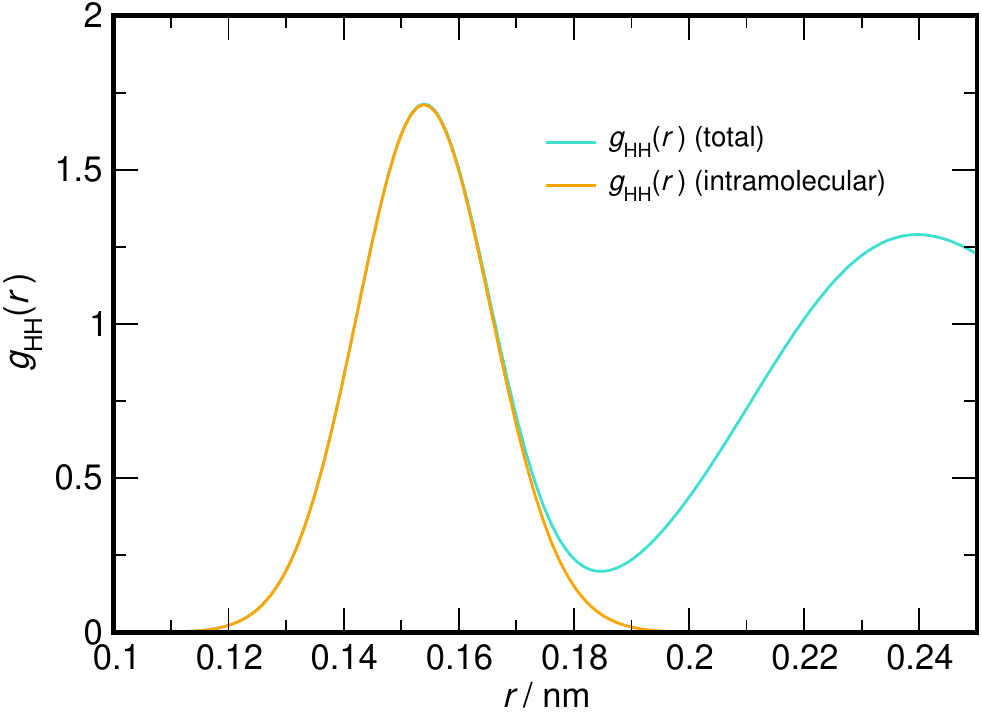}                     
        \caption{\label{fig:g2_water}
        	    Intramolecular and total H-H radial distribution function 
	    $g_\mathrm{HH}(r)$ 
	    for $T\!=\!298\,\mbox{K}$ as obtained from CCMD simulations
	    of H$_2$O.
	    The average intramolecular H-H distance is obtained to be
	    $\langle r_\mathrm{HH}^{-3}\rangle^{-1/3}\!=\!154.1\,\mbox{pm}$.}
\end{figure}
\begin{figure}
        \includegraphics[width=0.4\textwidth]{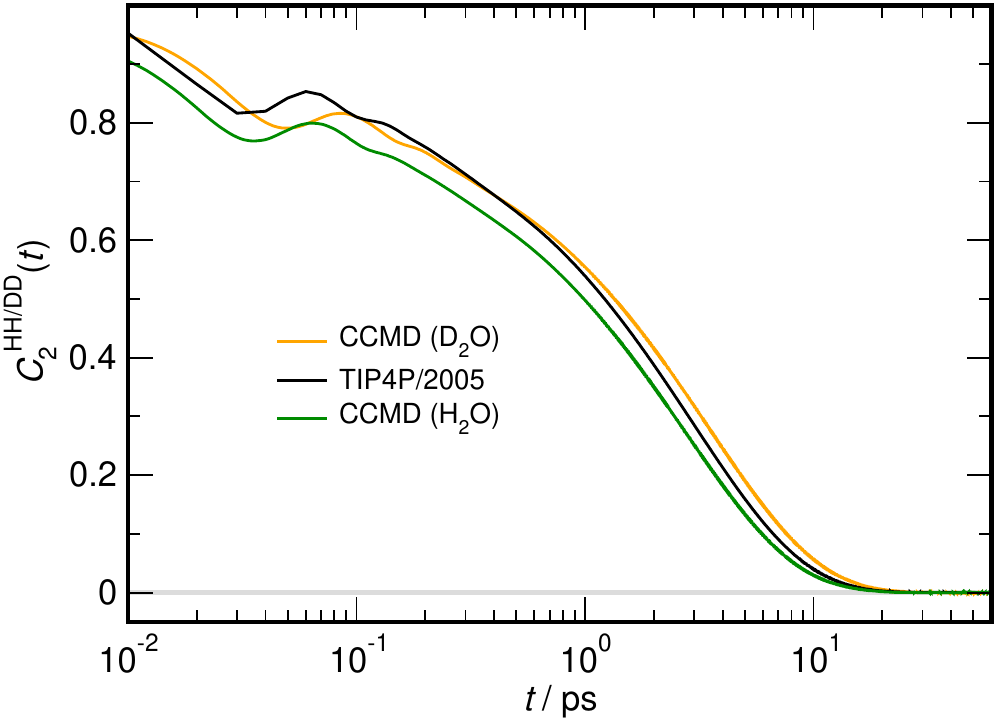}                     
        \caption{\label{fig:chh}
        	    Reorientational correlation function of the H-H-vector
	    $C_2^\mathrm{HH}(t)$. The corresponding 
	    reorientational correlation times are:
	    $\tau_2^\mathrm{HH}\!=\!(2.16 \pm 0.02)\,\mbox{ps}$ (CCMD H$_2$O),
		$\tau_2^\mathrm{HH}\!=\!(2.48 \pm 0.01)\,\mbox{ps}$ (TIP4P/2005),
		and
		$\tau_2^\mathrm{DD}\!=\!(2.80 \pm 0.04)\,\mbox{ps}$ (CCMD D$_2$O).}
\end{figure}
\begin{figure}
        \includegraphics[width=0.4\textwidth]{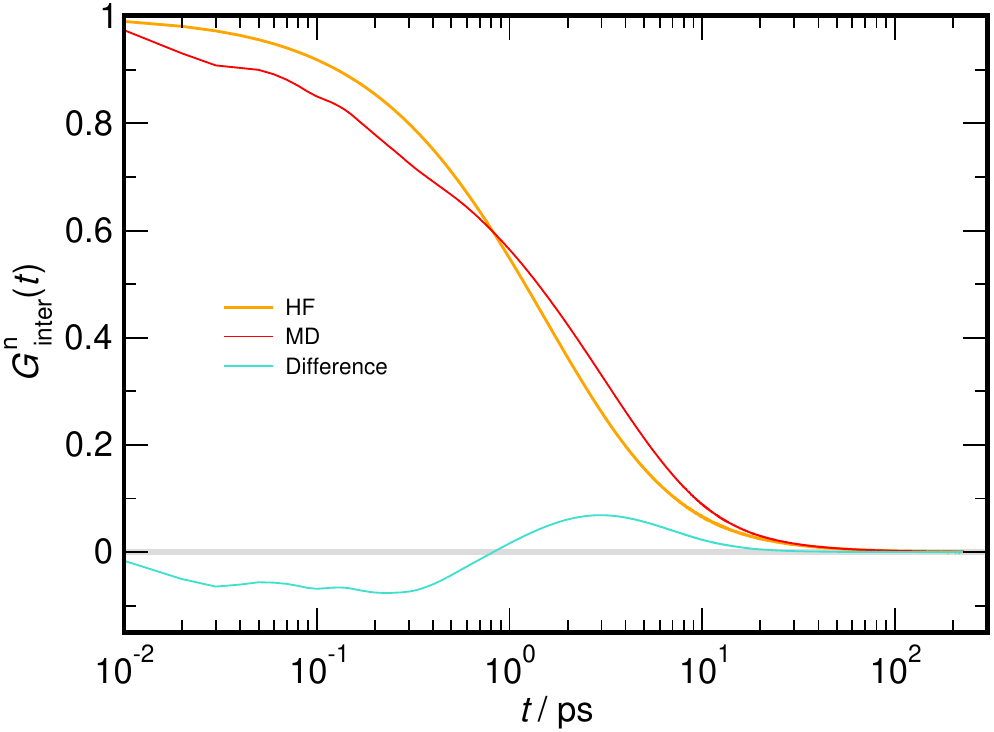}        
        \caption{\label{fig:g2_inter}
         Normalized intermolecular 
         dipole-dipole correlation function
                $G^\mathrm{n}_\mathrm{inter}(t)$
        computed for water at $298\,\mbox{K}$.
	     Intermolecular dynamics from classical MD simulations
	     of TIP4P/2005 water.
	     Solid red line: $G^\mathrm{n,MD}_\mathrm{inter}(t)$.
         Solid orange line:
	    $G^\mathrm{n,HF}_\mathrm{inter}(t)$ computed using data shown in
	    \tablename\ I of Ref. \cite{paschek_2024,paschek_2024cor}.
	    Solid turquoise line: difference function
	    	    $\Delta G^\mathrm{n}_\mathrm{inter}(t)=G^\mathrm{n,MD}_\mathrm{inter}(t)-G^\mathrm{n,HF}_\mathrm{inter}(t)$.
	    }
\end{figure}
\begin{figure}
        \includegraphics[width=0.4\textwidth]{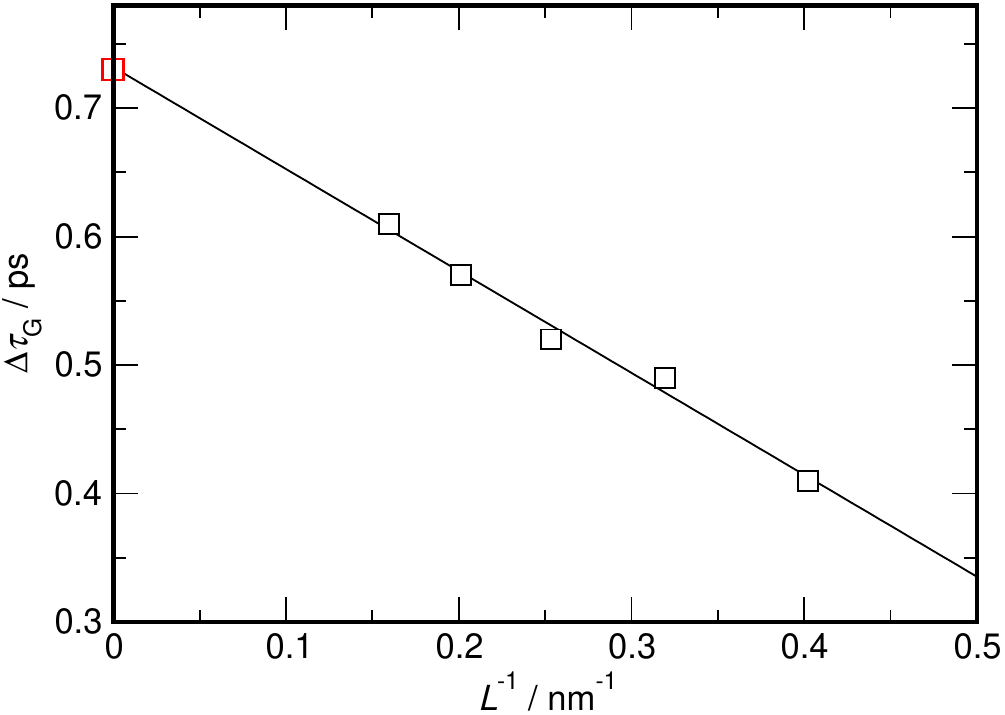}        
        \caption{\label{fig:deltataug}
        Scaling of the computed
         intermolecular
         $\Delta \tau_\mathrm{G}$ for
         TIP4P/2005 water at $298\,\mbox{K}$
         given in \tablename\ I of Ref. \cite{paschek_2024,paschek_2024cor}
         as a function of the inverse
         box length $L^{-1}$, analogous to the scaling of the translational diffusion
         coefficient suggested by Yeh and Hummer.\cite{yeh_2004}
         The extrapolated value for $L\!\rightarrow\!\infty$ is indicated in red.
	    }
\end{figure}
\begin{figure*}
        \includegraphics[width=0.4\textwidth]{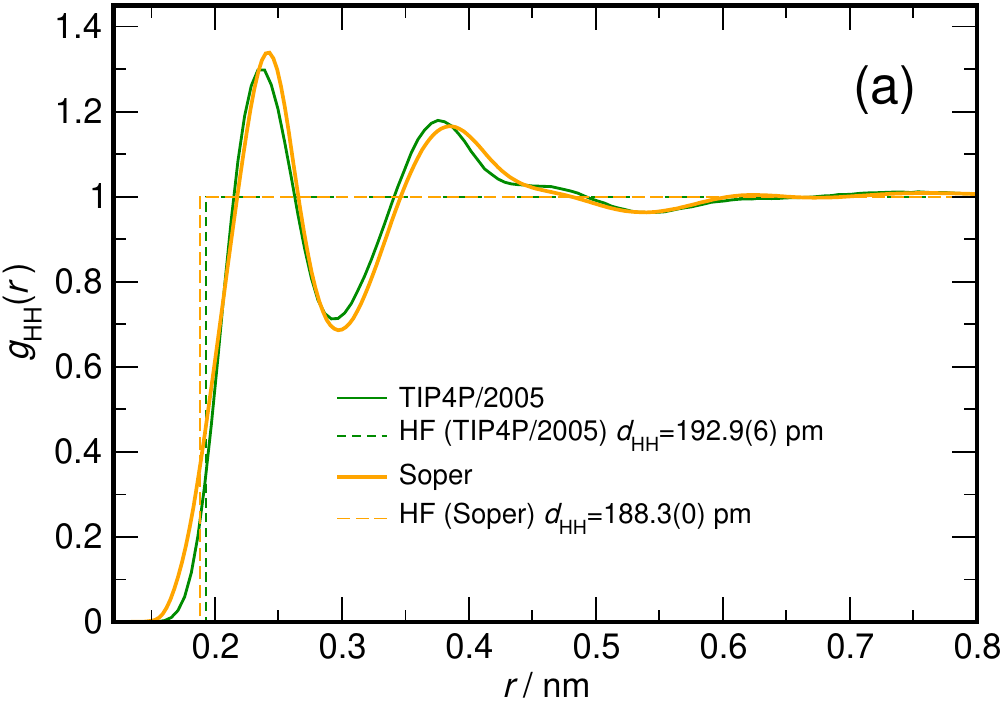}     
        \includegraphics[width=0.4\textwidth]{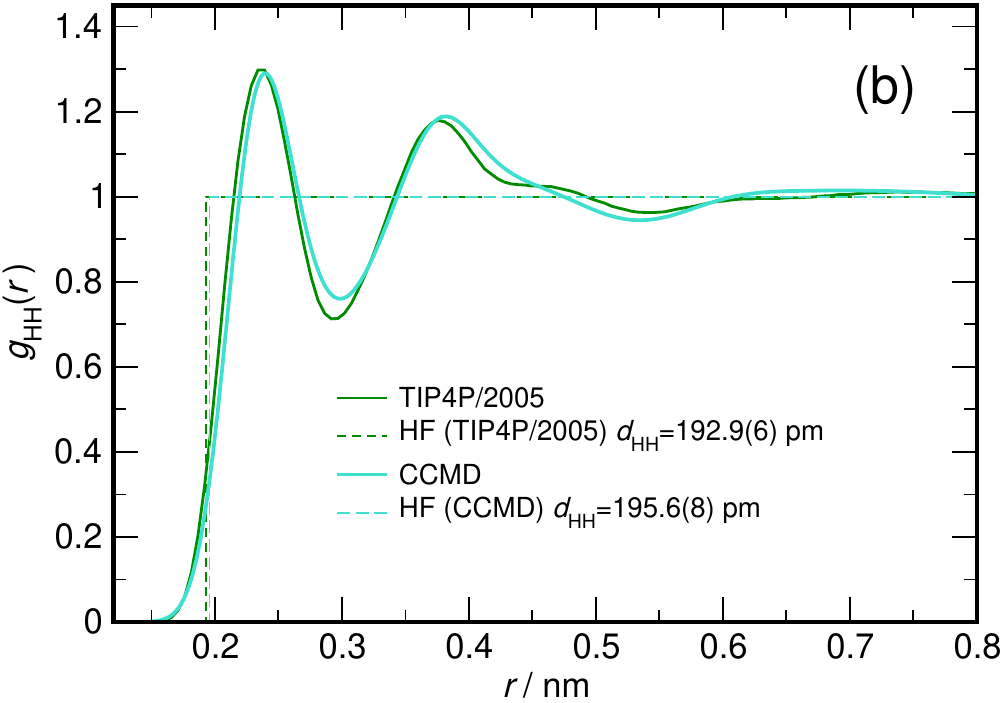}             
        \caption{\label{fig:gofr_hh}
	    Intermolecular H-H radial distribution function 
	    $g_\mathrm{HH}(r)$ 
	    for $T\!=\!298\,\mbox{K}$ 
	    in addition to the corresponding
	    step-like $g_\mathrm{HH}(r)$ according to
	    the Hwang and Freed theory with 
	    DCAs of $d_\mathrm{HH}$ as indicated.
	    a) A comparison of the radial distribution functions obtained
	    for the TIP4P/2005 model \cite{paschek_2024,paschek_2024cor} and from neutron scattering
	    data according to Soper.\cite{soper_2013}
	    b) A comparison of the radial distribution functions obtained
	    for the TIP4P/2005 model \cite{paschek_2024,paschek_2024cor} and from CCMD simulations
	    of H$_2$O.}
\end{figure*}

As discussed in section \ref{sec:theory}, the normalized
intramolecular dipole-dipole correlation function $G^\mathrm{n,MD}_\mathrm{intra}(t)$ 
is dominated by the 
reorientational correlation function $C_2^\mathrm{HH}(t)$ of
the intramolecular H-H vector as defined by Equation \ref{eq:reorhh}.
For the case of
the rigid TIP4P/2005 model,
$G^\mathrm{n,MD}_\mathrm{intra}(t)$ 
is perfectly identical with
$C_2^\mathrm{HH}(t)$,
since the distance of the H-H vector is being
kept fixed during the simulation.
This is, however, not the
case for the CCMD simulations, where no constraints with respect to
bond distances and bond angles exist. Here the length of
the H-H connecting vector will be mostly affected by intramolecular
bond bending vibrational motions. Bond bending
and reorientational motions
are, however, well de-correlated since
the H-H vibrations are about two orders of magnitude faster
than the reorientational motions, 
such that
\begin{equation}
\label{eq:intra_approx1}
G^\mathrm{n,MD}_\mathrm{intra}(t)\approx
\frac{\langle
r_\mathrm{HH}^{-3}(0)
r_\mathrm{HH}^{-3}(t)
\rangle}
{\langle r_\mathrm{HH}^{-6}\rangle}\cdot
C_2^\mathrm{HH}(t)\;.
\end{equation} 
A consequence of the fast de-correlation of the oscillatory motion
is that
the distance-dependent
time correlation function is nearly instantly quenched, which leads to
\begin{equation}
\label{eq:intra_approx2}
G^\mathrm{n,MD}_\mathrm{intra}(t)\approx
\frac{\langle
r_\mathrm{HH}^{-3}
\rangle^2}
{\langle r_\mathrm{HH}^{-6}\rangle}\cdot
C_2^\mathrm{HH}(t)\;.
\end{equation} 
This would correspond to employing
$C_2^\mathrm{HH}(t)$ instead of $G^\mathrm{n,MD}_\mathrm{intra}(t)$
to determine
$J^\mathrm{n,MD}_\mathrm{intra}(\omega)$
in Equation \ref{eq:R1_intra} while using
for the term $1/r_\mathrm{HH}^6$ in Equation \ref{eq:R1_intra}
\begin{equation}
\label{eq:r3-weighting}
\frac{1}{r_\mathrm{HH}^6} =
\langle
r_\mathrm{HH}^{-3}
\rangle^2\;.
\end{equation}

In Refs. \cite{paschek_2024,paschek_2024cor} we have employed
the intramolecular H-H distance of the 
TIP4P/2005 model 
with
$r_\mathrm{HH}= 151.4\,\mbox{pm}$
leading to an intramolecular relaxation rate
of  $R_{1,\mathrm{intra}}(\nu_\mathrm{H})\!=\!0.1759\,\mbox{s}^{-1}$,
which is practically frequency independent for resonance frequencies
$\nu_\mathrm{H}\!\leq\!400\,\mbox{MHz}$.
In 2021 Faux et al.\cite{faux_2021} revisited
the reorientational dynamics of water
and argued in favour of using the L\'evy rotor model
as the most appropriate description. They also collected information about the 
intramolecular H-H distances from a variety of experimental 
sources and concluded a distance of
$\langle r_\mathrm{HH}^{-3}\rangle^{-1/3}\!=\!(154.5\pm 0.7)\,\mbox{pm}$ as a consensus distance for
liquid water in the condensed phase (see supporting information of Ref.\cite{faux_2021}).
This larger distance would by itself lead to
a reduction of the intramolecular relaxation rate
by about $11.45\,\%$ compared to the value given above.\cite{paschek_2024,paschek_2024cor} Here, we have computed
the average intramolecular H-H distance
according to Equation \ref{eq:r3-weighting}
from the intramolecular pair correlation function 
obtained from our CCMD simulations of H$_2$O
shown
in \figurename\ \ref{fig:g2_water},
yielding a value of 
$\langle r_\mathrm{HH}^{-3}\rangle^{-1/3}\!=\!154.1\,\mbox{pm}$, which is in excellent agreement with
the estimate of Faux et al.

In \figurename\ \ref{fig:chh} we show 
$C_2^\mathrm{HH}(t)$
reorientational correlation functions
obtained for H$_2$O and D$_2$O obtained from our CCMD simulations in addition to
the reorientational correlation function obtained from the
purely classical
TIP4P/2005 simulations. 
We would like to emphasize that
there is a great similarity between the librational
short-time features of the TIP4P/2005 model and the 
CCMD data for H$_2$O. The CCMD derived function for H$_2$O is, however, situated significantly
below the classical TIP4P/2005 data, suggesting that the correlation
function is quenched due to the quantum delocalization of 
the hydrogen nuclei. This is supported by the CCMD data for D$_2$O
also shown in \figurename\ \ref{fig:chh}, where
this quenching effect is less pronounced. 
For D$_2$O, the larger reduced mass also leads to a visibly enlarged librational 
frequency as indicated
by its increased oscillation period. The corresponding reorientational correlation time for
H$_2$O with
$\tau_2^\mathrm{HH}\!=\!(2.162 \pm 0.017)\,\mbox{ps}$,
is accordingly reduced by about $12.8\,\%$ compared to the
$\tau_2^\mathrm{HH}\!=\!(2.48 \pm 0.01)\,\mbox{ps}$ of the
TIP4P/2005 model, while the value for 
D$_2$O with
$\tau_2^\mathrm{DD}\!=\!(2.80 \pm 0.04)\,\mbox{ps}$
is larger than both.
By using
$\tau_2^\mathrm{HH}\approx\tau_\mathrm{G}$ and 
$r_\mathrm{HH}\!=\!154.1\,\mbox{pm}$ according to 
Equation
\ref{eq:R1_intra}, we compute 
an intramolecular relaxation rate of
$\lim_{\omega\rightarrow 0}R_{1,\mathrm{intra}}(\omega)
\!=\!0.1380\,\mbox{s}^{-1}$
in the extreme narrowing limit for H$_2$O from CCMD simulations,
which is $21.6\,\%$ smaller than the value obtained for the
TIP4P/2005 model.

The various dynamical modes within a liquid are intricately
related to the overall fluidity of the liquid matrix. 
According to hydrodynamic theory the
translational and rotational dynamics are related to the viscosity of
the fluid via the Stokes-Einstein (SE) and Stokes-Einstein-Debye (SED) relations,
respectively. Combining both SE and SED relations lead to the
expression \cite{kawasaki_2019}
\begin{equation}
\label{eq:balance}
D_0\times \tau_l =
\frac{4}{3}\cdot
\frac{1}{l(l+1)}\cdot
R_\mathrm{h}^2\,,
\end{equation}
where $D_0$ is the translational self-diffusion coefficient, 
$\tau_l$ is the reorientational correlation time determined from a correlation function
of a $l$-th order Legendre-polynomial, while $R_\mathrm{h}$ represents the hydrodynamic radius.
The observation that the corresponding expression 
$\tau_l/\tau_l'=l'(l'+1)/(l(l+1))$ is not fulfilled for 
certain molecular systems has been
regarded as an indication of a ``breakdown'' of the Debye model.\cite{turton_2014,koeddermann_2008}
Moreover,
given the important role that hydrogen bond exchange plays in the detailed mechanism
of both water reorientation \cite{laage_2006,laage_2008} and
water diffusion \cite{gomez_2022}, it might even seem surprising how well
the computed hydrodynamic radius agrees with the actual diameter of 
a water molecule \cite{herrero_2022} (at least for water under ambient conditions).
In essence, however, Equation \ref{eq:balance} describes a balance between reorientational
and translational dynamics that is often
surprisingly close to what has been predicted according to the
hydrodynamic theory.\cite{turton_2014}
Note that
delocalization of the H nuclei might affect translational and
reorientational modes in liquid water differently and will thus
 have an effect on this balance.
To investigate this effect quantitatively, 
we have computed the term 
$D_0\times \tau^\mathrm{HH}_2$ for all three systems under consideration.
Here we notice
a systematic variation
with
$D_0\times \tau^\mathrm{HH}_2\!=\!0.506\,\mbox{\AA}^2$
for H$_2$O from CCMD simulations,
$D_0\times \tau^\mathrm{DD}_2\!=\!0.521\,\mbox{\AA}^2$
for D$_2$O from CCMD simulations, and 
$D_0\times \tau^\mathrm{HH}_2\!=\!0.570\,\mbox{\AA}^2$
for the classical TIP4P/2005 simulations.
It is important to realize that  the inclusion of
nuclear quantum delocalization in the simulation
affects the reorientational dynamics of water more strongly than
the translational motion, leading to a decrease of
$D_0\times \tau^\mathrm{HH}_2$ for 
H$_2$O as compared to D$_2$O. Moreover, both values are significantly
smaller than 
$D_0\times \tau^\mathrm{HH}_2\!=\!0.570\,\mbox{\AA}^2$
computed here for the purely classical simulation of TIP4P/2005 water.
We would like to point out that, in addition to the 
intramolecular H-H distance, a correct representation of this 
$D_0\times \tau^\mathrm{HH}_2$
balance is essential for providing an accurate estimate for the  
intramolecular relaxation rate $R_{1,\mathrm{intra}}$.
The diffusion coefficient for H$_2$O from CCMD simulations is
found to be slightly larger than the experimental self-diffusion
coefficient.
To account for the effect on this balance, we suggest the following procedure
for determining $\tau_\mathrm{G}$
\begin{equation}
\tau_\mathrm{G}\approx \tau_2^\mathrm{HH}\times
\frac{D_0(\mathrm{CCMD},\mathrm{H}_2\mathrm{O})}{D_0(\mathrm{expt.})}\;.
\end{equation}
When including this small modification, 
we compute as the best estimate for the
 intramolecular relaxation rate 
according to Equation
\ref{eq:R1_intra} of
$\lim_{\omega\rightarrow 0}R_{1,\mathrm{intra}}(\omega)
\!=\!0.1404\,\mbox{s}^{-1}$
in the extreme narrowing limit for H$_2$O.
Computed values for various frequencies $\nu_\mathrm{H}$ 
using the spectral density function
$J^\mathrm{n,MD}_\mathrm{intra}(\omega)$ obtained from 
CCMD simulations
can be found
in \tablename\ \ref{tab:md4}.
\begin{figure*}
        \includegraphics[width=0.4\textwidth]{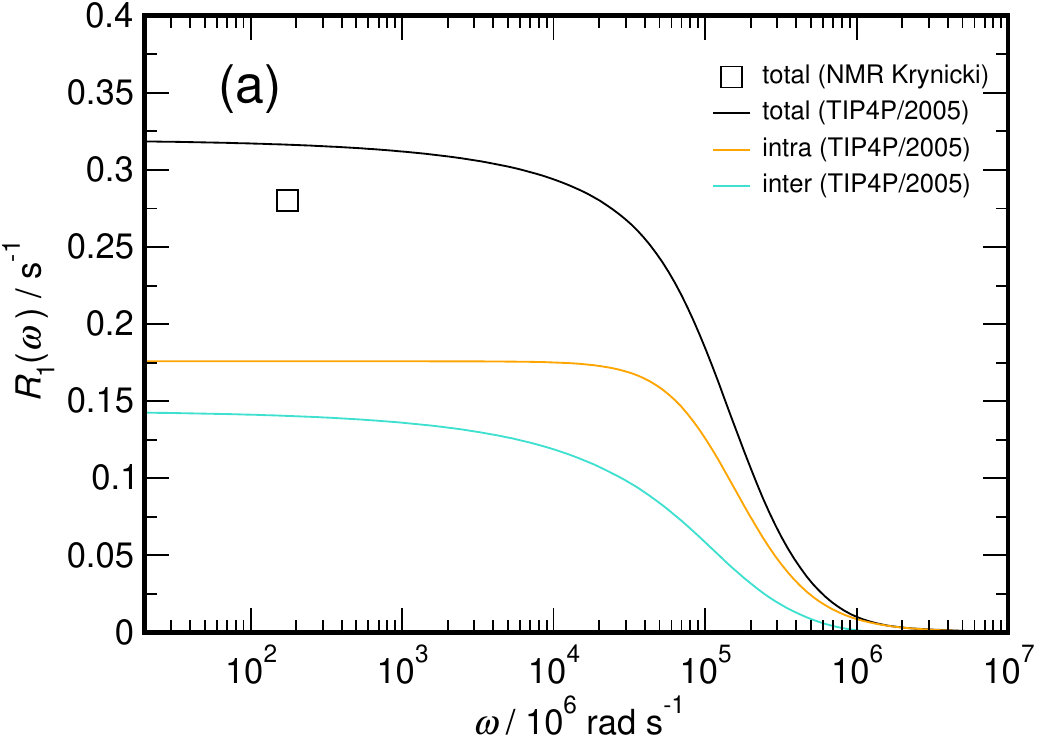}        
        \includegraphics[width=0.4\textwidth]{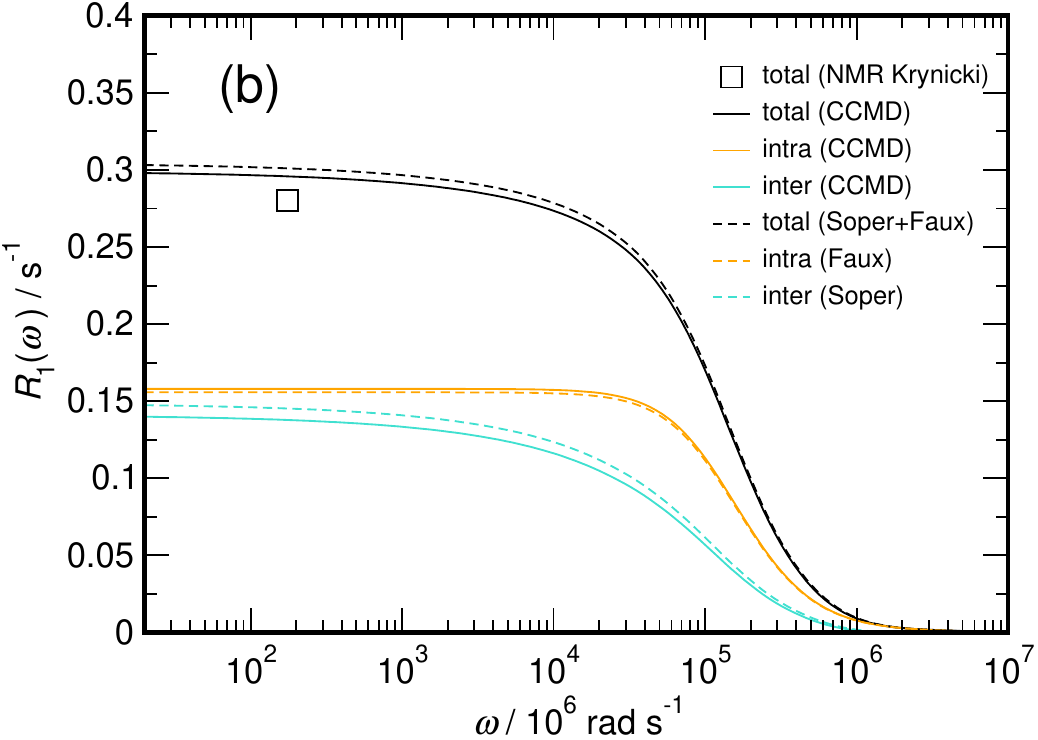}        
        \caption{\label{fig:R1_all}
Frequency-dependent $^1$H NMR relaxation rate $R_1(\omega)$ of water at 298 K at a density of 
$0.997\,\mbox{g}\,\mbox{cm}^{-3}$ a) based on TIP4P/2005 model data
as in Ref. \cite{paschek_2024,paschek_2024cor} using $D'=2D_0$ with
$D_0\!=\!2.30\times 10^{-9}\,\mbox{m}^2\,\mbox{s}^{-1}$ and
b) employing refined structural parameters 
for $d_\mathrm{HH}$ and $\langle r^{-3}_\mathrm{HH}\rangle^{-1/3}$
either based on CCMD
simulations or the experimental data
of Soper \cite{soper_2013} and Faux et al.\cite{faux_2021}. 
	    }
\end{figure*}
	
\begin{figure*}
        \includegraphics[width=0.4\textwidth]{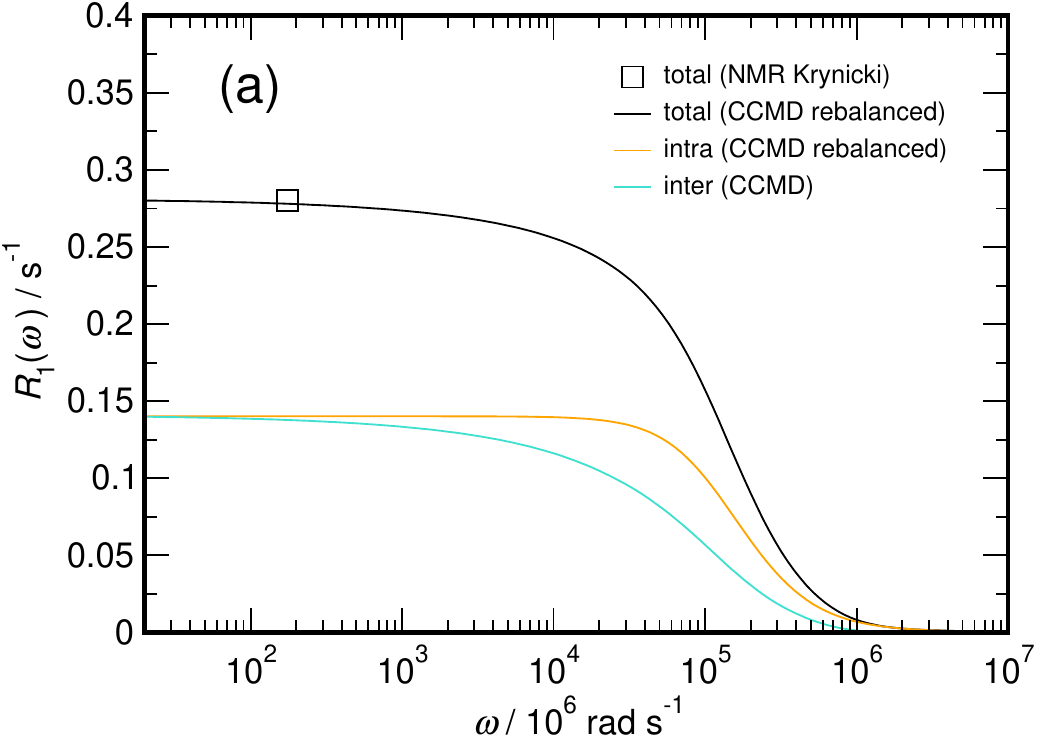}        
        \includegraphics[width=0.4\textwidth]{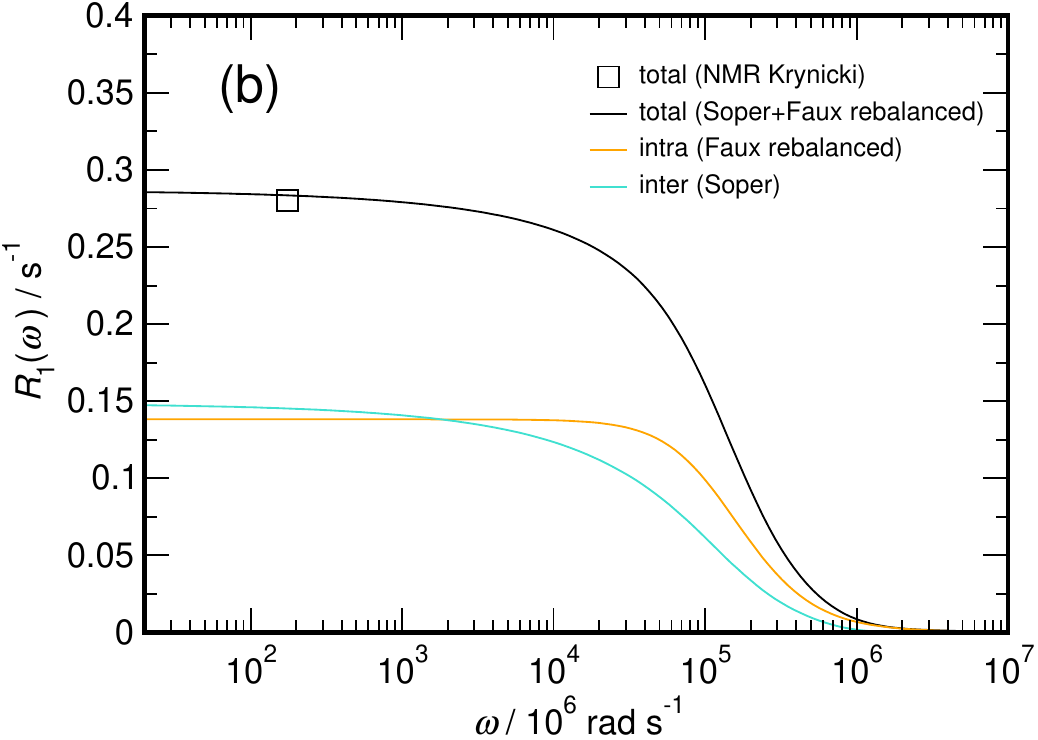}        
        \caption{\label{fig:R1_rebalanced}
Refined frequency dependent $^1$H NMR relaxation rate $R_1(\omega)$ of water at 298 K at a density of 
$0.997\,\mbox{g}\,\mbox{cm}^{-3}$
computed using the intramolecular spectral
densities $J^\mathrm{n}_\mathrm{intra}(\omega)$ obtained from CCMD 
scaled to match the experimental balance
of $D_0\times\tau_\mathrm{HH}$
a) employing the refined structural parameters from CCMD and
b) 
employing the refined structural parameters based on 
the experimental data.}
\end{figure*}

\subsection{Intermolecular Relaxation}

For long times $G^\mathrm{n}_\mathrm{inter}(t)$
shows a $t^{-3/2}$ scaling behavior and the curve determined
from MD simulation asymptotically approaches that of the Hwang and Freed model.
In Ref. \cite{paschek_2024,paschek_2024cor} we have shown that
for times $t\geq t_\mathrm{tr}$ both curves are practically indistinguishable.
A log-linear representation of the data, including the
difference function
\begin{equation}
\Delta G^\mathrm{n}_\mathrm{inter}(t)=G^\mathrm{n,MD}_\mathrm{inter}(t)-G^\mathrm{n,HF}_\mathrm{inter}(t)
\end{equation}
is shown in \figurename\ \ref{fig:g2_inter}. Note that the difference
function is negative up to a time of about $1\,\mbox{ps}$, then turns positive until 
it asymptotically approaches zero.
These negative and positive contributions
are due to a deviation of the dynamics of the $^1$H nuclei in liquid water
from a continuous micro-step diffusion mechanism
assumed in the Hwang-Freed model.
They are due to both, the librational motions of the water molecules, and
the jump-like reorientational dynamics of
water molecules discussed in detail by D. Laage and J.T. Hynes.\cite{laage_2006,laage_2008}
The intermolecular 
correlation time $\tau_\mathrm{G}$ can be computed 
as an integral over $G^\mathrm{n,MD}_\mathrm{inter}(t)$, which can hence 
be splitted into two terms
according to
\begin{equation}
\tau_\mathrm{G} = \tau_\mathrm{G,HF} + \Delta\tau_\mathrm{G}
\end{equation}
with $\tau_\mathrm{G,HF}=4/9\cdot d^2_\mathrm{HH}/D'$
following Equation \ref{eq:taug}. Here $\Delta\tau_\mathrm{G}$
can be computed comfortably via numerical integration of
\begin{equation}
\Delta\tau_\mathrm{G} \approx \int\limits_0^{t_\mathrm{tr}} 
\Delta G^\mathrm{n}_\mathrm{inter}(t) \;dt
\end{equation}
due to the short-time nature of $\Delta G^\mathrm{n}_\mathrm{inter}(t)$.
To compute the intermolecular
spectral density from MD simulation, we use
\begin{equation}
J^\mathrm{n,MD}_\mathrm{inter}(\omega)
=
J^\mathrm{n,HF}_\mathrm{inter}(\omega)
+
\Delta J^\mathrm{n}_\mathrm{inter}(\omega)
\end{equation}
with
\begin{equation}
\label{eq:deltaj_md}
\Delta J^\mathrm{n}_\mathrm{inter}(\omega)
\approx
\int\limits_0^{t_\mathrm{tr}} 
\Delta G^\mathrm{n}_\mathrm{inter}(t)
\cos (\omega t) \;dt\;.
\end{equation}
Here the integration 
in Equation \ref{eq:deltaj_md}
is performed numerically 
employing the trapezoidal rule
up to the time $t_\mathrm{tr}$, where
both functions $G^\mathrm{n,MD}_\mathrm{inter}(t)$
and $G^\mathrm{n,HF}_\mathrm{inter}(t)$ are deemed indistinguishable.
To properly predict $J^\mathrm{n,MD}_\mathrm{inter}(\omega)$ for systems
with $L\rightarrow\infty$, we employ the system size independent
self-diffusion coefficient $D_0$ for computing $J^\mathrm{n,HF}_\mathrm{inter}(\omega)$.
In addition, we use
\begin{equation}
\Delta J^\mathrm{n}_\mathrm{inter}(\omega)
\approx
\frac{\Delta\tau_{\mathrm{G},N\rightarrow\infty}}{\Delta\tau_{\mathrm{G},N=8192}}
\cdot
\Delta J^\mathrm{n}_{\mathrm{inter},N=8192}(\omega)
\end{equation}
to predict the behavior
of $\Delta J^\mathrm{n}_\mathrm{inter}(\omega)$ for infinite system sizes.
Here $\Delta J^\mathrm{n}_{\mathrm{inter},N=8192}(\omega)$ is the 
difference function computed for a system containing 8192 water molecules, and
$\Delta\tau_{\mathrm{G},N=8192}=\Delta J^\mathrm{n,MD}_{\mathrm{inter},N=8192}(0)$ 
and $\Delta\tau_{\mathrm{G},N\rightarrow\infty}=\Delta J^\mathrm{n,MD}_{\mathrm{inter},N\rightarrow\infty}(0)$ is the
corresponding correlation time predicted for an infinite system size via
extrapolation. The values used here 
are shown in \figurename\ \ref{fig:deltataug} as a function of system size and
are given in
Table I of Ref. \cite{paschek_2024} with
$\Delta\tau_{\mathrm{G},N\rightarrow\infty}/\Delta\tau_{\mathrm{G},N=8192}\!=\!1.1967$
for TIP4P/2005 water at $T\!=\!298\,\mbox{K}$.
The intermolecular
$^1$H NMR relaxation rate
following Equation \ref{eq:relax}
is finally computed via
\begin{eqnarray}
\label{eq:R1_inter}
R_{1,\mathrm{inter}}(\omega)
&=&
\gamma_\mathrm{H}^4\hbar^2 \cdot\frac{3}{4}\cdot\left(\frac{\mu_0}{4\pi}\right)^2\cdot
\frac{8\pi}{15}\cdot\frac{\rho_\mathrm{H}}{d_\mathrm{HH}^3}\times\\
&&\left\{
J^\mathrm{n,MD}_\mathrm{inter}(\omega)
+
4\,J^\mathrm{n,MD}_\mathrm{inter}(2\omega)
\right\}\;.
\nonumber
\end{eqnarray}
Here $\rho_\mathrm{H}$ is representing the number density of
the $^1$H nuclei in liquid water.

The effect of the intermolecular water structure on the relaxation rate is captured
by the DCA $d_\mathrm{HH}$. For the case of the 
TIP4P/2005 model for water, we have determined this value to be
$d_\mathrm{HH}\!=\!192.96\,\mbox{pm}$. Here we have used simulations containing
8192 water molecules to compute the 
H-H radial distribution functions shown in \figurename\ \ref{fig:gofr_hh}.
To determine  $d_\mathrm{HH}$ we have applied Equation \ref{eq:dhh-2}
while
the $r^{-6}$ weighted integral over the pair distribution function
was computed according to Equation \ref{eq:dhh-3}. 
The obtained value for 
$d_\mathrm{HH}\!=\!192.96\,\mbox{pm}$ 
includes a long-range correction as described by
Equation \ref{eq:dhh-3} and is found to be independent of the chosen
cutoff-radius for $R_\mathrm{c}\geq 0.9\,\mbox{nm}$
in Equation \ref{eq:dhh-3}.
We have also computed corresponding DCAs
from the experimental pair distribution data according to
Soper \cite{soper_2013}
 shown in \figurename\ \ref{fig:gofr_hh}a, leading to
 $d_\mathrm{HH}\!=\!188.30\,\mbox{pm}$,
 and the CCMD dataset for H$_2$O shown 
 in \figurename\ \ref{fig:gofr_hh}b, leading to
 $d_\mathrm{HH}\!=\!195.68\,\mbox{pm}$.
 When using the same TIP4P/2005 derived
$\Delta J^\mathrm{n}_\mathrm{inter}(\omega)$ functions for all three datasets,
we determine
relaxation rates at $\nu_\mathrm{H}\!=\!28\,\mbox{MHz}$ of
$R_{1,\mathrm{inter}}(\nu_\mathrm{H})\!=\!0.1405\,\mbox{s}^{-1}$ 
for the TIP4P/2005 model,
$R_{1,\mathrm{inter}}(\nu_\mathrm{H})\!=\!0.1453\,\mbox{s}^{-1}$ 
for the experimental structure and
$R_{1,\mathrm{inter}}(\nu_\mathrm{H})\!=\!0.1378\,\mbox{s}^{-1}$ 
for the CCMD dataset for H$_2$O.
Note that the deviation with respect to the relaxation rate of
the TIP4P/2005 model are in both cases less than $2\,\%$.
Hence, the intermolecular relaxation rate $R_{1,\mathrm{inter}}$
is found to be not very sensitive with respect to small changes of
$d_\mathrm{HH}$. 
The reason for this can be explained by the Hwang Freed theory 
which shows that the DCA 
 $d_\mathrm{HH}$ affects both the
prefactor and the correlation time
$\tau_\mathrm{G}^\mathrm{HF}\!=\!(4/9)\cdot d^2_\mathrm{HH}/D'$ such that
$R_{1,\mathrm{inter}}\!\propto\!1/d_\mathrm{HH}$.
Given the insensitivity of the computed intermolecular
relaxation rate with respect to structural information collected
from different sources, and the fact that diffusion coefficients obtained 
for 
both TIP4P/2005 and CCMD simulations agree with the experimental value,
we conclude that the computed intermolecular relaxation rates
reported here can be considered a rather robust estimate.

\subsection{Computing the Frequency-Dependent NMR Relaxation of $^1$H Nuclei in Liquid Water}

In \figurename\ \ref{fig:R1_all}a 
we have plotted the 
frequency-dependent
inter- and intramolecular contributions to the
$^1$H NMR  relaxation rate according to Equations \ref{eq:R1_inter}
and \ref{eq:R1_intra} for the TIP4P/2005 model
as discussed in Ref.\cite{paschek_2024}.
By using the information about $d_\mathrm{HH}$ and 
$r_\mathrm{HH}$ from CCMD simulations and experimental
data we arrive at the frequency-dependent relaxation rates
shown in
\figurename\ \ref{fig:R1_all}b.
Note that refining just the structural parameters already significantly
improves the agreement with the experimental data. This effect
has to be mostly attributed to the improved intramolecular H-H distance.

In \figurename\ \ref{fig:R1_rebalanced}a and
\ref{fig:R1_rebalanced}b the rebalanced intramolecular spectral density 
obtained from CCMD simulations is used in addition to the modified
structural parameters $d_\mathrm{HH}$ and 
$r_\mathrm{HH}$ obtained from 
CCMD simulations (\figurename\ \ref{fig:R1_rebalanced}a) and
from experimental data (\figurename\ \ref{fig:R1_rebalanced}b).
The data computed from CCMD simulations agree within the experimental
error with the experimental relaxation rate. 
Moreover, our analysis suggests that both the intra-
and intermolecular contribution to the relaxation rate are almost identical
in magnitude.

Finally, in Table \ref{tab:md4}, we report the computed
relaxation rates for H$_2$O shown in \figurename\ \ref{fig:R1_rebalanced}a
for frequencies accessible via
state-of-the-art NMR hardware. In the discussed frequency range the intramolecular
relaxation rate is almost frequency independent, even at an 
$^1$H resonance frequency of $1.2\,\mbox{GHz}$ the relaxation rate deviates by only about
$0.2$ per  cent. Due to the long-time nature of
the intermolecular dipolar correlation function 
following a $t^{-3/2}$ power law, the intermolecular relaxation rate shows a much stronger frequency dependence with a drop of about $15$ per cent at $1.2\,\mbox{GHz}$.
\begin{table}
\caption{\label{tab:md4}
Intermolecular, intramolecular, and total
dipolar $^1$H NMR relaxation rates as a function of the
frequency $\nu=\omega/(2\pi)$
employing refined structural parameters based on CCMD
data while
matching the balance
of $D_0\times\tau_\mathrm{HH}$ according to
 the CCMD simulations of H$_2$O water; see text.
The experimental relaxation rate
was obtained to be
$R_1(\nu_\mathrm{H})\!=\!(0.280\pm 0.003)\,\mbox{s}^{-1}$
at a frequency of 
$\nu_\mathrm{H}\!=\!28\,\mbox{MHz}$.\cite{Krynicki_1966}}
\setlength{\tabcolsep}{0.155cm}
        \centering              
\begin{tabular}{cccc}
\hline\hline\\[-0.2em]
$\nu_\mathrm{H}/\mbox{MHz}$ & 
$R_{1,\mathrm{inter}}(\nu_\mathrm{H})/\mbox{s}^{-1}$ &
$R_{1,\mathrm{intra}}(\nu_\mathrm{H})/\mbox{s}^{-1}$ &
$R_{1}(\nu_\mathrm{H})/\mbox{s}^{-1}$ 
\\\hline\\[-0.6em]
0    & 0.1410  & 0.1404 & 0.2814   \\
28   & 0.1378  & 0.1404 & 0.2782   \\
50   & 0.1368  & 0.1404 & 0.2772   \\
200  & 0.1325  & 0.1404 & 0.2729   \\
400  & 0.1289  & 0.1404 & 0.2692   \\
800  & 0.1238  & 0.1403 & 0.2641   \\
1000 & 0.1216  & 0.1402 & 0.2618   \\
1200 & 0.1197  & 0.1401 & 0.2598    \\[0.6em]\hline\hline
\end{tabular}
\end{table}

\section{Conclusions}

We have applied a recently introduced computational framework aimed at 
reliably
determining the frequency-dependent
intermolecular NMR dipole-dipole relaxation
to compute the $^1$H NMR relaxation rate of liquid
water at ambient conditions. We have achieved this by
using information from
highly accurate
Coupled Cluster Molecular Dynamics (CCMD) 
trajectories for liquid
H$_2$O and D$_2$O
\cite{Daru2022Coupled},
which employ a high-dimensional neural network potential
generated at CCSD(T) accuracy
while accounting for the
effect of quantum delocalization of the nuclei
by employing
path integral simulations.
%
We observe a close-to-perfect agreement with experimental
$^1$H NMR relaxation data
if both structural and dynamical information from  CCMD 
trajectories
are
taken into account, while also
including a re-balancing of the rotational and translational dynamics,
according to
product of the self-diffusion coefficient and
the reorientational correlation time of the H-H vector $D_0\times\tau_\mathrm{HH}$.
The inclusion of nuclear quantum effects 
significantly reduces
the computed intramolecular contribution to the
$^1$H NMR relaxation rate.
In addition, our analysis suggests that the intermolecular and intramolecular
contribution to the $^1$H NMR relaxation rate are almost similar in magnitude, unlike
to what has been predicted earlier from classical MD simulations. 
As a conclusion, we can state that by employing state-of-the-art 
condensed matter simulation techniques, a quantitative 
understanding of the $^1$H NMR relaxation in liquid water is within reach.

%
%
%

\section*{Acknowledgements}

DP would like to thank David W. Sawyer for sharing his
knowledge about the early history of 
NMR relaxation experiments on water and for
his helpful comments.
AS thanks the
Deutsche Forschungsgemeinschaft
(DFG, German Research Foundation) 
under Grant STR-1626/2-1 (Project 459405854) for supporting this work.
RL and AMCT acknowledge the Marie 383 Sk\l{}odowska-Curie 
Actions Doctoral Network FC Relax 384 (HORIZON-MSCA-DN-2021, 10107275) for funding.
NS, HF, and DM acknowledge funding by the DFG under Germany's
Excellence Strategy~-- EXC~2033~-- 390677874~-- RESOLV.

\section*{Author Declarations}
 
\subsection*{Conflict of Interest}

The authors have no conflicts to disclose.
 
\section*{Data Availability Statement}

The data that support the findings of this study are available from the corresponding author upon reasonable request.


\bibliography{all}

\end{document}